\documentclass[aps,prb,twocolumn,10pt,superscriptaddress,notitlepage]{revtex4-1}
\newcommand{\papertitle}{Electrical 2$\pi$ phase control of infrared light in a 350\,nm footprint using graphene plasmons}

\usepackage[utf8]{inputenc}
\usepackage{graphicx}
\renewcommand{\figurename}{Figure}
\usepackage{graphicx}
\usepackage{amsmath}
\usepackage{natbib}
\usepackage{float}
\usepackage{tikz}
\usepackage{siunitx}
\usepackage{dcolumn}
\usepackage{lmodern}
\usepackage[T1]{fontenc}
\usepackage[unicode=true,
	    colorlinks=true,
	    linkcolor=blue,
	    citecolor=blue,
	    urlcolor=blue]{hyperref} 
\usepackage{sansmath,caption}
\newcommand{\newtextbar}[1][3]{\scalebox{#1}[1]{\textbar}}
\DeclareCaptionLabelSeparator{vertline}{\textbf{~\newtextbar}}
\usepackage[font=sf, labelfont={sf,bf},size=small,justification=RaggedRight,textfont={sf,sansmath},labelsep=vertline]{caption}
\usepackage{ulem}
\usepackage[percent]{overpic}

\newcommand{\Vps}{\ensuremath{V_{\mathrm{PS}}}}
\newcommand{\kp}{\ensuremath{k_{\mathrm{pl}}}}
\newcommand{\pwl}{\ensuremath{\lambda_{\mathrm{pl}}}}

\newcommand{\pinvloss}{\ensuremath{\gamma_{\mathrm{p}}^{-1}}}

\newcommand{\ns}{\ensuremath{n_\mathrm{s}}}
\renewcommand{\Re}{\operatorname{Re}}

\begin{document}

\title{\Large\textsf{\papertitle}}


\author{Achim Woessner}
\thanks{These authors contributed equally}
\affiliation{\footnotesize ICFO - Institut de Ciencies Fotoniques, The Barcelona Institute of Science and Technology, 08860 Castelldefels (Barcelona), Spain}
\author{Yuanda Gao}
\thanks{These authors contributed equally}
\affiliation{\footnotesize Department of Mechanical Engineering, Columbia University, New York, NY 10027, USA}
\author{Iacopo Torre}
\affiliation{\footnotesize NEST, Scuola Normale Superiore, I-56126 Pisa, Italy}
\affiliation{\footnotesize Istituto Italiano di Tecnologia, Graphene Labs, Via Morego 30, I-16163 Genova, Italy}
\author{Mark B. Lundeberg}
\affiliation{\footnotesize ICFO - Institut de Ciencies Fotoniques, The Barcelona Institute of Science and Technology, 08860 Castelldefels (Barcelona), Spain}
\author{Cheng Tan}
\affiliation{\footnotesize Department of Mechanical Engineering, Columbia University, New York, NY 10027, USA}
\author{Kenji Watanabe}
\affiliation{\footnotesize National Institute for Materials Science, 1-1 Namiki, Tsukuba 305-0044, Japan}
\author{Takashi Taniguchi}
\affiliation{\footnotesize National Institute for Materials Science, 1-1 Namiki, Tsukuba 305-0044, Japan}
\author{Rainer Hillenbrand}
\affiliation{\footnotesize CIC nanoGUNE and UPV/EHU, 20018 Donostia-San Sebastian, Spain}
\affiliation{\footnotesize IKERBASQUE, Basque Foundation for Science, 48011 Bilbao, Spain}
\author{James Hone}
\affiliation{\footnotesize Department of Mechanical Engineering, Columbia University, New York, NY 10027, USA}
\author{Marco Polini}
\affiliation{\footnotesize Istituto Italiano di Tecnologia, Graphene Labs, Via Morego 30, I-16163 Genova, Italy}
\author{Frank H.L. Koppens}
\email{frank.koppens@icfo.eu}
\affiliation{\footnotesize ICFO - Institut de Ciencies Fotoniques, The Barcelona Institute of Science and Technology, 08860 Castelldefels (Barcelona), Spain}
\affiliation{\footnotesize ICREA – Institució Catalana de Recerça i Estudis Avancats, 08010  Barcelona, Spain}

\maketitle

\begin{figure}[t]
\includegraphics{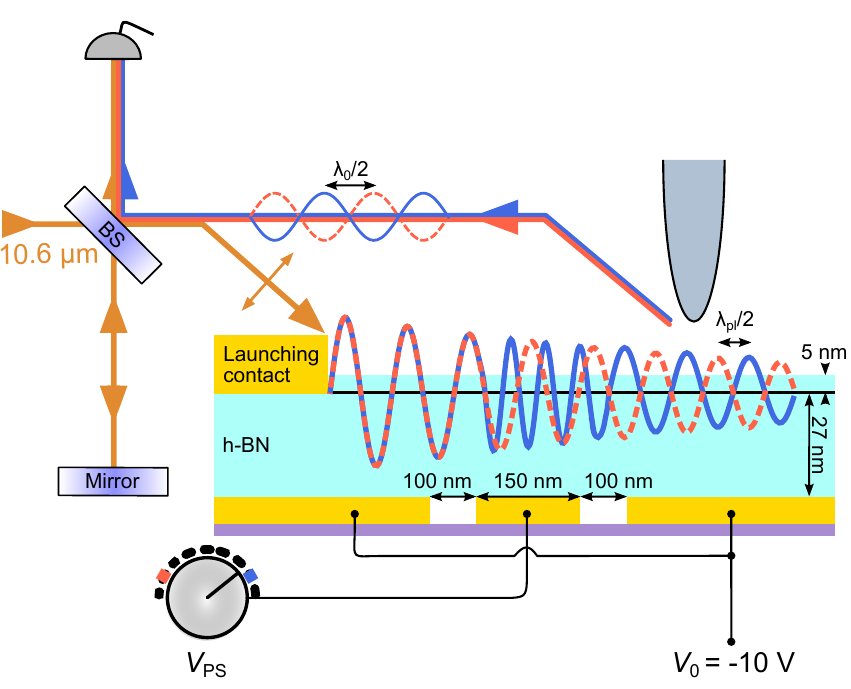}

\caption{
\label{fig1}
\textbf{Device sketch and measurement principle.}
The three local gates are used to independently tune the carrier concentration in the different graphene regions.
The phase shifter voltage $\Vps$ is tuned and during all experiments $V_0 =  -10$\,V is kept constant. 
The gate thickness is 15\,nm.
The plasmons propagate from the launching contact over the phase shifter region and are ultimately scattered into far-field light using a metallized AFM tip, and subsequently interfered with the incoming light.
}\end{figure}

\begin{figure}[t]
\includegraphics{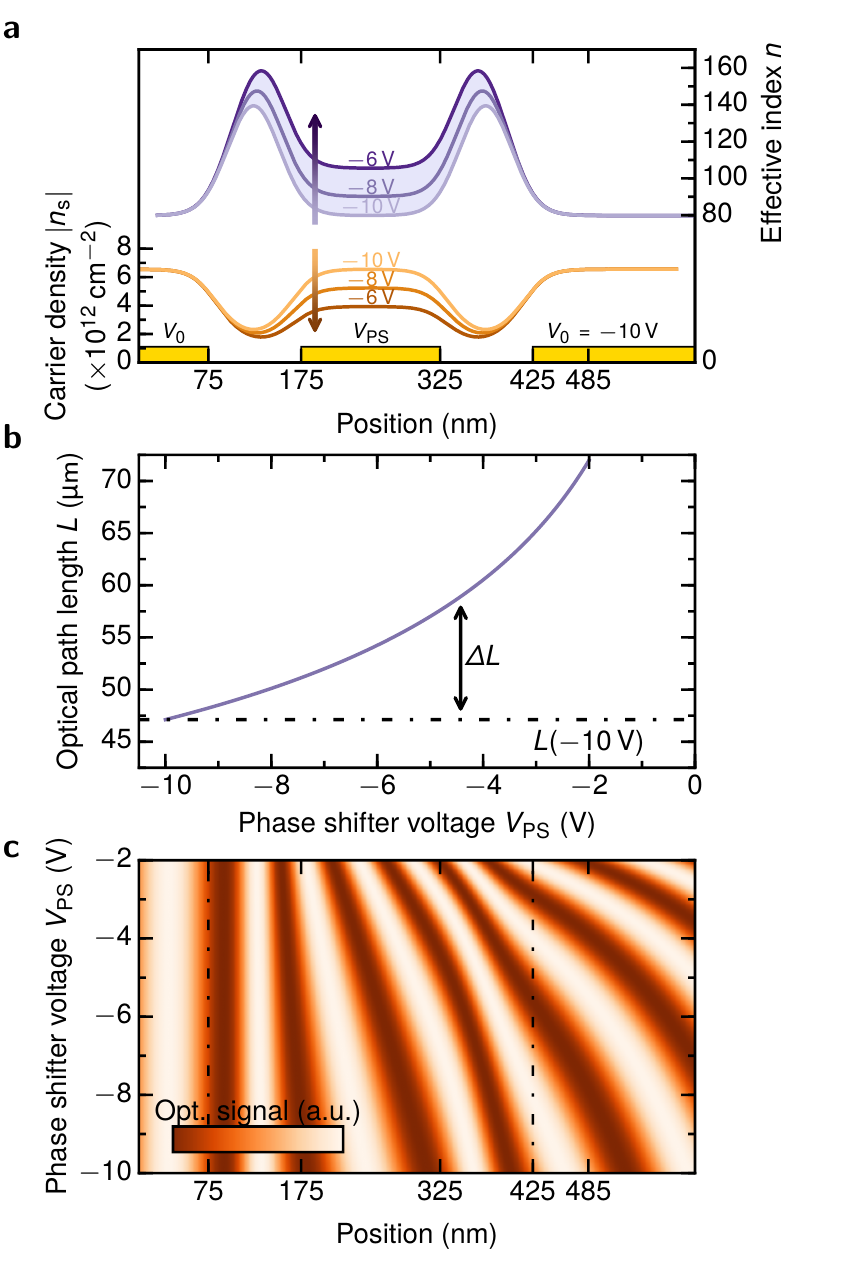}

\caption{
\label{fig2}
\textbf{Working principle.}
\textbf{a},
Simulated spatial carrier density profile and the corresponding spatial profile of the plasmon effective index $n$, for a range of $\Vps$.
Varying $\Vps$ leads to different optical path lengths $L$.
The shaded area indicates the optical path difference $\Delta L = \SI{7}{\micro\metre}$ between $\Vps=-10$\,V and $\Vps=-6$\,V, equivalent to a phase shift of $\Delta \phi = 4/3\pi$ at $\lambda_0 = \SI{10.6}{\micro\metre}$.
The position of the local gates is indicated at the bottom.
\textbf{b},
Calculated optical path length $L$ from 0\,nm to 485\,nm for a changing $\Vps$ and a constant $V_0 = -10$\,V.
The optical path difference for the different configurations is indicated as $\Delta L$.
\textbf{c},
Plasmons propagate from the left and start accumulating a different phase for different $\Vps$ inside the phase shifting region (indicated by the two dashed lines) due to their different optical path lengths $L(\Vps)$.
}\end{figure}

\begin{figure}[t]

\includegraphics{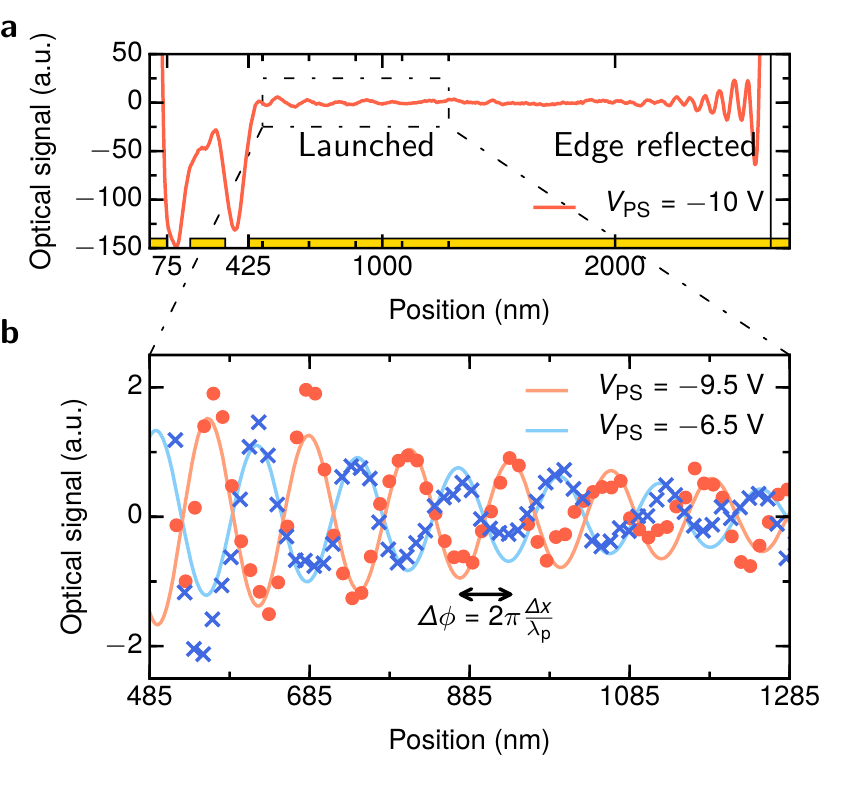}
\caption{
\label{fig3}
\textbf{Measurement of full device and a phase shift of $\pi$.}
\textbf{a},
Optical signal of the full device with the launching contact at 0\,nm and the position of all the gates indicated.
For clarity a polynomial background was subtracted. 
The edge of the graphene is indicated as vertical solid line at 2670\,nm.
\textbf{b},
Measurement (data points) and fits (solid lines) of the absolute value of the fourth harmonic of the optical signal with a smoothed optical signal background subtracted from both.
The phase shift between $\Vps = -9.5$\,V and $\Vps = -6.5$\,V is $(1.04\pm 0.11)\pi$.
}\end{figure}

\begin{figure}[t]
\centering
\includegraphics{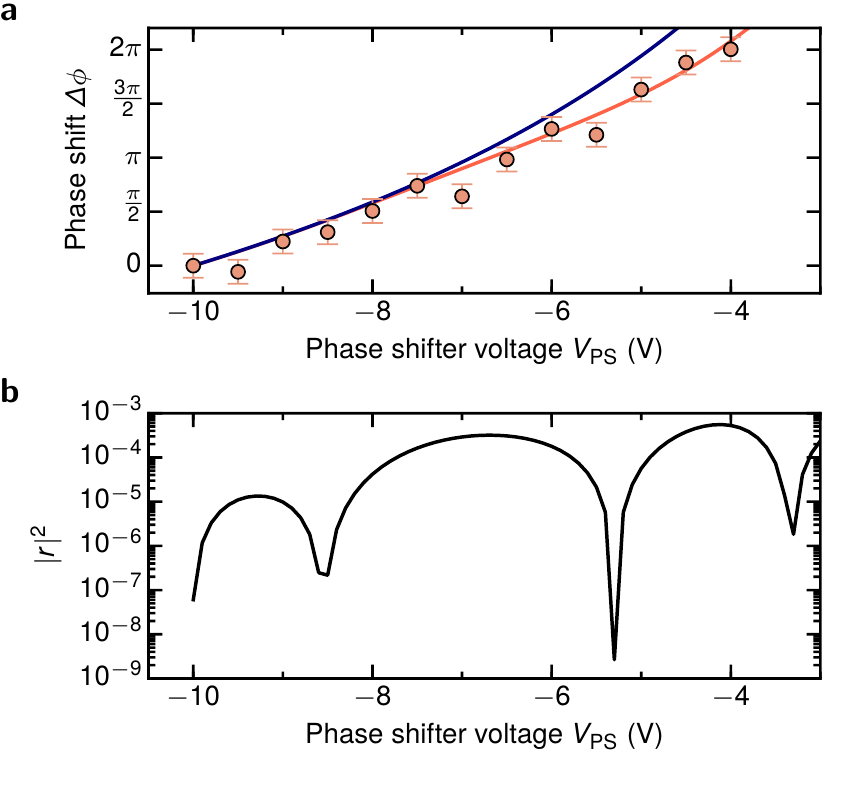}
\caption{
\label{fig4}
\textbf{Phase shift $\Delta \phi$ and reflection magnitude.}
\textbf{a} Calculated and measured phase shift with respect to the phase of the transmitted plasmon at a gate voltage of $-10$\,V applied to both the large gates and the phase shifter gate.
The blue line is the phase shift calculated from $\Delta  \phi = k_0\Delta L$ using the calculated optical path difference $\Delta L$.
The red line is the phase shift calculated from Lippmann-Schwinger theory (see supplement).
The error bar length of $\pm 0.11\pi$ is the standard deviation of the extracted phase of the launched plasmon for 12 measurements with $\Vps = V_0=-10$\,V.
\textbf{b} Calculated reflection magnitude from the phase shifting device for a plane wave plasmon coming in from the left.
}\end{figure}

\textsf{\textbf{
Modulating the amplitude and phase of light is at the heart of many applications such as wavefront shaping,\cite{Dickey} transformation optics,\cite{Pendry2006,Vakil2011} phased arrays,\cite{Sun2013} modulators\cite{Yu2014b} and sensors.\cite{Takeda1982}
Performing this task with high efficiency and small footprint is a formidable challenge.\cite{Reed2010a,Liu2015}
Metasurfaces\cite{Kildishev2013,Yu2014b} and plasmonics\cite{Dionne2009} are promising
, but metals exhibit weak electro-optic effects. 
Two-dimensional materials, such as graphene, have shown great performance as modulators with small drive voltages.\cite{Mohsin2015,Sun2016}
Here we show a graphene plasmonic phase modulator which is capable of tuning the phase between 0 and 2$\boldsymbol{\pi}$ \textit{in situ}. 
With a footprint of 350\,nm it is more than 30 times smaller than the 10.6\,$\boldsymbol{\mu}$m free space wavelength.
The modulation is achieved by spatially controlling the plasmon phase velocity 
in a device where the spatial carrier density profile is tunable.
We provide a scattering theory for plasmons propagating through spatial density profiles.
This work constitutes a first step towards two-dimensional transformation optics\cite{Vakil2011} for ultra-compact modulators\cite{Reed2010a} and biosensing.\cite{Rodrigo2015a}
}}

Graphene plasmons are a versatile tool for integrated photonics and nano-optoelectronics as they provide extreme sub-wavelength confinement of light\cite{Fei2012,Chen2012,Woessner2014} while still offering a long lifetime approaching 1\,ps.\cite{Woessner2014}
The graphene plasmon phase velocity (and thus wavelength) is \textit{in situ} tunable, and can be varied spatially, making it a unique platform for for transformation optics in two dimensions.\cite{Vakil2011,Wang2014d}
The full spatial and dynamic control of the plasmon velocity profile makes completely new device concepts possible. 
This includes on-chip interferometers and tunable metamaterials such as phased arrays for a full dynamical beam manipulation.\cite{Sun2013}
This full control has thus far remained a great challenge and the plasmon propagation was mainly controlled by physical features in the graphene.\cite{Fei2013,Chen2013a,Alonso-Gonzalez2014}

Here, we manipulate for the first time the spatial profile of the plasmon phase velocity actively employing local metal gates, achieving \textit{in situ} control of the plasmon wavelength, as sketched in Fig.~\ref{fig1}. 
A plasmon (launched by scattering light on a gold edge) that propagates trough the tunable spatial carrier density profile picks up a phase, that is transferred to the photon after scattering of a metallized atomic force microscopy probe tip. 
In this way, we are able to continuously tune the phase shift of the plasmon and the outcoming photons from 0 to $2\pi$, and experimentally measure this phase in the far-field with an interferometer. \cite{Phillips2000,Balistreri2000}

We present a simple model based on the optical path length of light to explain our observations and to provide guidelines for designing such graphene plasmonic phase modulators.
Moreover we rigorously calculate the expected phase shift and reflections using a Lippmann-Schwinger scattering theory approach.
Indeed our work is the first time for graphene plasmons that such an approach has been used.
This strongly improved understanding both theoretically and experimentally of plasmonic phase-control is relevant for the future development of \textit{in situ} tunable metasurfaces,\cite{Li2015a} modulators and holds great promise for novel sensor concepts.

Our device is based on a heterostructure of graphene encapsulated between two layers of hexagonal boron nitride (h-BN),\cite{Woessner2014} which serves to preserve the lifetime of the graphene plasmons. 
The device was assembled by the polymer-free van der Waals assembling technique\cite{Wang2013} and then transferred onto 15\,nm thin AuPd local gates.
The so-called ``phase shifter gate'' has a length of 150\,nm and the gaps to the other gates are 100\,nm each.
Using these local gates it is possible to spatially control the carrier density profile.
A gold contact is connected by electrical side contact,\cite{Wang2013} allowing for gating of the graphene.

A CO$_2$ laser with a free space wavelength of $\lambda_0 = \SI{10.6}{\micro\metre}$ is focussed onto this elongated gold contact with a sharp edge.
The polarization is perpendicular to the edge as sketched in Fig.~\ref{fig1}.
This edge provides the necessary momentum matching between far-field photons and plasmons and thus the light is partially converted into graphene plasmons propagating away from the edge as a plane wave in the electrostatically doped graphene.\cite{Huber2005,Alonso-Gonzalez2014} 

The effect of the plasmon propagation trough this spatial carrier density profile is depicted in Fig. ~\ref{fig2}a.
The plasmon effective index $n=c/v_\mathrm{ph}$, with $v_\mathrm{ph}$ the plasmon phase velocity and $c$ the speed of light, is related to the gate-induced charge density: $v_\mathrm{ph} \propto \pwl \propto\sqrt{\ns}$. \cite{Woessner2014} 
The voltage on the phase shifter gate $\Vps$ controls $n$ in the graphene above this gate.
After the plasmons propagate through this velocity-tunable (and thus phase-shifting) region their optical path length (Fig.~\ref{fig2}b) can be expressed as
\begin{equation}
\label{OPL}
L(\Vps)=\int_{a}^{b} n(x,\Vps)dx
\end{equation}
where $x$ is the position and $a=0$\,nm and $b=485$\,nm are the start and end points of the phase-shifting region respectively,\cite{BornWolf1999} 
where the carrier density profile does not change any more when $\Vps$ is changed.

The carrier density profile is simulated by the electrostatic Laplace equation taking into account the full anisotropic DC permittivity of h-BN as well as the spatial gate profile.\cite{Liu2013d}
For each $\Vps$ the carrier density profile is simulated and $n(x,\Vps)$ is then calculated taking into account the full dielectric environment (see supplement).
The carrier-density-dependent graphene conductivity is calculated using the zero-temperature non-local random phase approximation.\cite{Polini2008}
As the out-of-plane decay length of the plasmons is much smaller than the thickness of the bottom h-BN the spatial profile of the metal gates does not need to be taken into account.

Due to the different optical path lengths for different $\Vps$ the plasmons accumulate a different phase $\phi$ after propagating through the phase shifter region.
The phase difference $\Delta \phi$ between different optical path lengths can be estimated using the optical path difference, 
\begin{equation}
\label{eq:Delta phi}
\Delta \phi(\Vps) = k_0 \Delta L=k_0\left(L(\Vps)-L(-10\,{\mathrm{V}})\right).
\end{equation}
In Fig.~\ref{fig2}c we show how the optical path difference $\Delta L$ translates into a different phase for different $\Vps$ as they exit the phase shifter region.
The wavelength reduction in the phase shifter region is clear.

In order to experimentally observe the phase shift between different $\Vps$ we use a metallized atomic force microscopy probe tip to scan across the sample.
The plasmons are scattered from the probe tip and then recorded as the out-scattered light signal in the far-field by scattering-type scanning near-field optical microscopy (s-SNOM) (Fig.~\ref{fig1}).
The measurement is performed in homodyne configuration, where the light passes trough interferometric arms with a constant length.
While scanning the tip across the sample we observe a plasmon fringe pattern, as shown in Fig.~\ref{fig3}a, with the full plasmon wavelength $\pwl$ launched by the sharp metal edge on the left.
This fringe pattern is measured because the incident electric field interferes with the plasmon field.\cite{Huber2005,Alonso-Gonzalez2014}
We do not observe any plasmons reflected by the gold contact.
We also observe a fringe pattern with $\pwl/2$ at the graphene edge due to edge reflected plasmons interfering with the plasmons launched by the tip.\cite{Fei2012,Chen2012,Woessner2014}
The extracted plasmon wavelength at a gate voltage of $V_0=-10$\,V is $\pwl = 126\pm 4$\,nm for both launched and reflected plasmons (Fig.~\ref{fig3}a), in good agreement with the theoretically calculated 128\,nm.\cite{Woessner2014}
The drop of the optical signal at the gap position is mainly explained by the change in dielectric environment below the surface.
Gold is known to have a high optical signal however the missing gold in the gaps leads to a drop of the optical signal.

By changing $\Vps$ we observe a change in position of the launched plasmon fringes to the right of the phase shifter region (Fig.~\ref{fig3}b).
The phase shift can be extracted from the data using the simple relationship $\Delta \phi = \kp\Delta x$, with $\kp=2\pi/\pwl$. 
We find that the wavelength of the plasmons to the right of the phase shifting region is independent of $\Vps$.
This confirms that that the carrier density is well defined for the region where we extract the light phase, and that the phase-shift only occurs in the phase-shifting region for positions from 100 to 350\,nm.

In order to extract $\phi$ for different $\Vps$ we remove the background optical signal by subtracting the smoothed optical signal both from data and fitting function. 
We then fit launched and edge reflected fringes using $\Re(re^{i\phi}e^{ik x})$, where $\phi$ is the phase we are interested in and $r$ a real valued amplitude.
The plasmon wavevector $k=\kp + k_\mathrm{i}i = 2\pi/\SI{126}{\nano\metre} - \SI{1.5i}{\micro\metre}^{-1}$ is fixed in order to capture the phase fully in $\phi$. 
The extracted inverse damping ratio $\pinvloss=\kp/k_\mathrm{i}\sim 30$ is in accordance with previous studies on high quality graphene encapsulated in h-BN at room temperature.\cite{Woessner2014}
In order to compensate for small drifts in the interferometric arm length which can lead to a change in the fringe position, we measure the phase of the launched plasmons relative to the phase of the edge reflected plasmons.
The edge reflected plasmons are not influenced by $\Vps$ as the graphene edge is $\sim$\SI{2.7}{\micro\metre} away (and thus more than the plasmon decay length)  from the phase shifter gate (Fig.~\ref{fig3}a).
The phase of the edge reflected plasmons is measured relative to the graphene edge from the simultaneously measured topography.\cite{Woessner2014}

In Fig.~\ref{fig4}a we show the measured phase shift $\Delta \phi$ relative to the plasmon phase measured at $\Vps=V_0=-10$\,V.
We observe a fully tunable phase shift from $0$ to $2\pi$ for a 6\,V range. 
The blue line is the phase shift calculated from the optical path length approximation (OPLA)\cite{BornWolf1999} using eq.~\ref{eq:Delta phi}, without fitting parameters.
A good agreement with experiments is reached for small $\Vps-V_0$ but stronger deviation is clearly visible for larger $\Vps-V_0$.

In order to quantitatively understand the behaviour of the phase shift, we employ a scattering approach that parallels Lippmann-Schwinger single-particle quantum scattering theory, which we termed Lippmann-Schwinger-Random Phase Approximation (LS-RPA). 
This approach reduces to the OPLA when the plasmon wavelength is much smaller than the length scale over which the carrier density changes. As for our device, these two lengthscales are comparable, the LS-RPA approach is required.
Numerical results obtained with the LS-RPA approach as shown in Fig.~\ref{fig4}a are in excellent agreement with experimental data.
The essence of the LS-RPA approach is the following:
We start by considering an inhomogeneous system where spatial variations of the carrier density profile are confined to a limited region of space.
The propagating plasmon modes outside this region define the incoming and outgoing plane wave modes.\cite{TorreTobepublished2016}
Complex transmission and reflection coefficients are evaluated from a long-wavelength expansion of the proper density-density linear response function.
This is the point where the Random Phase Approximation (RPA) for an inhomogeneous electron system \cite{TorreTobepublished2016} is carried out, to approximate the proper density response function with the non-interacting one. 
Keeping only the leading term in the long-wavelength limit, one recovers the local approximation,\cite{Garcia-Pomar2013} where only the local frequency-dependent conductivity function matters.
A more quantitative understanding of the data, however, requires the first non-local correction to the long-wavelength expansion (see supplement).

Using this approach we also calculate the expected optical rejection ratio $|r|^2$ for this device which is very small (Fig.~\ref{fig4}b), owing to the smooth change of the plasmon wavelength.
This makes this type of device an ideal candidate as a modulator as unwanted reflections are minimal.
Indeed we do not observe plasmons reflected by the split gate as there are no plasmon fringes with spacing $\pwl/2$ observed next to the phase shifting region.

One important figure-of-merit of a phase modulator is the half-wave voltage $V_\pi$, the voltage that needs to be applied to change the phase by $\pi$.
Smaller voltages means more efficient operation.
The voltage-length product $V_\pi l$ reached is \SI{2.5}{\volt \micro\metre}, more than an order of magnitude smaller than for plasmonic phase modulators based on other materials (\SI{60}{\volt \micro\metre}).\cite{Haffner2015}
It is also much smaller than for diffraction limited phase modulators based on pin-photodiodes (\SI{360}{\volt \micro\metre})\cite{Green2007} or pin-diodes with a photonic crystal (\SI{50}{\volt \micro\metre}).\cite{Hosseini2012}
Commercially used lithium niobate modulators have much larger voltage-length products on the order of \SI{}{\volt \centi\metre}.
By decreasing the thickness of the bottom h-BN $V_\pi$ can even be further reduced.
We have modelled our modulator using an equivalent circuit model (see supplement) and find that the  cutoff frequency can be up to $\sim 1$\,THz, but the real switching time will be limited be limited by other processes.
We remark that the coupling efficiency from light to plasmons for these devices is very low.
Improved matching schemes and resonant structures\cite{Li2015a} are required for the development of practical modulators or metasurfaces.

To conclude, we have implemented a novel approach for ultra-compact voltage-tunable plasmonic phase control by spatially controlling the graphene plasmon phase velocity.
By pushing towards higher carrier concentrations it could be possible to build such modulators in the technologically relevant telecommunication wavelength range around \SI{1.5}{\micro\metre}, as near-infrared plasmons have recently been observed.\cite{Wang2016}
The same type of device could also be used for amplitude modulation of light\cite{Gomez-Diaz2013} or on-chip sensing by plasmon interferometry.
By nano-structuring the phase shifting region, it would be possible to make phased arrays to control, steer and focus light \textit{in situ} on the nanoscale.\cite{Dickey,Sun2013}

\section*{\textsf{Methods}}
{\small

\subsection*{Measurement details.}
The s-SNOM used was a NeaSNOM from Neaspec GmbH, equipped with a CO$_2$ laser operated at \SI{10.6}{\micro\metre} with $\sim$ 20\,mW power.
The metallized probes were commercially-available atomic force microscopy probes with an apex radius of $\sim\SI{25}{\nano\metre}$.
The tip height was modulated at $\sim 250~\mathrm{kHz}$ with $\sim 100$\,nm amplitude.
The probe tip was electrically grounded.
The beam splitter used is a ZnSe window.
The measurements were all performed at ambient conditions and room temperature.
In order to have a background free signal the fourth harmonic of the scattered light signal was used.
The gate capacitance density is $7.29\times 10^{11}~e\,\mathrm{cm^{-2}\,V^{-1}}$, where $e$ is the elementary charge.
The h-BN DC permittivity is 6.7 in-plane and 3.56 out-of-plane and the permittivity at \SI{10.6}{\micro\metre} is $8.341+0.023i$ in-plane and $1.932+0.006i$ out-of-plane.\cite{Woessner2014}

\subsection*{Device fabrication.}
The local metal gates are defined using electron beam lithography and 15\,nm of a AuPd alloy is evaporated.
The graphene and h-BN are first exfoliated onto a freshly cleaned substrate.
The stack of h-BN/graphene/h-BN is then prepared using the polymer-free van der Waals assembling technique\cite{Wang2013} and is transferred onto the local gates.
Using electron beam lithography and reactive ion etching the graphene is shaped into a rectangle.
Subsequently the launching contact is defined using electron beam lithography and evaporation of 75\,nm Au.
} 


\section*{\textsf{Acknowledgements}}
{\small
We thank A.J. Huber, K.-J. Tielrooij, I. Epstein and W. Heni for fruitful discussions and D. Davydovskaya and G. Navickaite for assistance in the clean room.
Open source software was used (www.matplotlib.org, www.python.org, www.inkscape.org).
F.H.L.K. acknowledges financial support from the Spanish Ministry of Economy and Competitiveness, through the "Severo Ochoa" Programme for Centres of Excellence in R\&D (SEV-2015-0522), support by Fundacio Cellex Barcelona, the ERC starting grant (307806, CarbonLight), the Government of Catalonia trough the SGR grant (2014-SGR-1535), the Mineco grants Ramón y Cajal (RYC-2012-12281) and Plan Nacional (FIS2013-47161-P), and project GRASP (FP7-ICT-2013-613024-GRASP). 
F.H.L.K. and R.H. acknowledge support by the E.C. under Graphene Flagship (contract no. CNECT-ICT-604391).
Y.G. and J.H. acknowledge support from the US Office of Naval Research N00014-13-1-0662.
M.P. is extremely grateful for the financial support granted by ICFO during a visit in August 2016 and acknowledges Fondazione Istituto Italiano di Tecnologia.
K.W. and T.T. acknowledge support from the Elemental Strategy Initiative conducted by the MEXT, Japan and JSPS KAKENHI Grant Numbers JP26248061, JP15K21722 and JP25106006.
}
\section*{\textsf{Author contributions}}
{\small
A.W., M.B.L. and F.H.L.K. conceived the experiment.
A.W. performed the experiments and simulations, analysed the data and wrote the manuscript.
Y.G. and C.T. fabricated the devices.
I.T. and M.P. developed the LS-RPA.
M.B.L. helped with simulations and data analysis.
K.W. and T.T. synthesized the h-BN.
R.H., J.H. and F.H.L.K. supervised the work.
All authors contributed to the scientific discussion and manuscript revisions.
}
\section*{\textsf{Competing Financial Interests}}
{\small
R.H. is co-founder of Neaspec GmbH, a company producing scattering-type scanning near-field optical microscope systems such as the ones used in this study. 
All other authors declare no competing financial interests.
}

\section*{\textsf{Data Availability Statement}}
{\small
The data that support the plots within this paper and other findings of this study are available from the corresponding author upon reasonable request.
}

\renewcommand\refname{Supplementary references}
\def\bibsection{\section*{\refname}} 

\renewcommand{\figurename}{Supplementary~Figure}
\setcounter{equation}{0}
\setcounter{figure}{0}
\setcounter{table}{0}
\setcounter{page}{1}
\makeatletter
\renewcommand{\theequation}{S\arabic{equation}}
\renewcommand{\thefigure}{S\arabic{figure}}
\renewcommand{\thetable}{\arabic{table}}
\renewcommand{\thepage}{\roman{page}}
\renewcommand{\bibnumfmt}[1]{[#1]}
\renewcommand{\citenumfont}[1]{#1}


\begin{widetext}
\pagebreak
\maketitle

\textbf{\Large Supplementary Material: \papertitle}

\end{widetext}
\newcommand{\Eref}{\ensuremath{E_\mathrm{ref}}}
\newcommand{\Ezref}{\ensuremath{E_\mathrm{0,ref}}}
\newcommand{\ENF}{\ensuremath{E_\mathrm{NF}}}
\newcommand{\EzNF}{\ensuremath{E_\mathrm{0,NF}}}
\newcommand{\EBG}{\ensuremath{E_\mathrm{BG}}}
\newcommand{\EzBG}{\ensuremath{E_\mathrm{0,BG}}}
\newcommand{\xref}{\ensuremath{x_\mathrm{ref}}}
\newcommand{\xtip}{\ensuremath{x_\mathrm{tip}}}

\section{Device geometry}

\begin{figure}[h]
\centering
\includegraphics{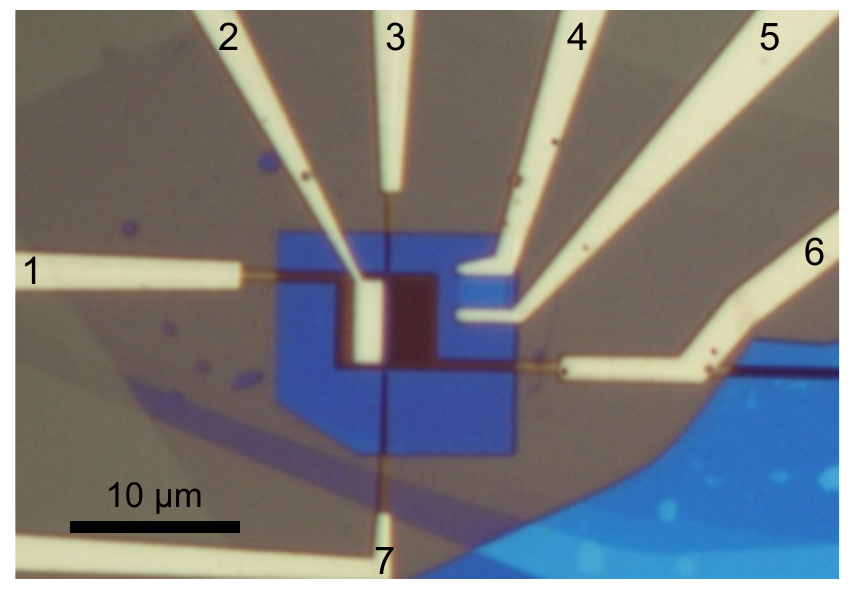}
\caption{
\label{fig-supp-device}
\textbf{Microscope image of the device used for the study.}
The distance between the launching gate and the graphene edge is \SI{2.6}{\micro\metre}.
The large local gates can be gated by the contacts 1 and 6 on the left and right, respectively.
Contacts 3 and 7 control the phase shifter gate voltage.
Contact 2 is the side-contacted launching contact pad.
Terminals 4 and 5 are contacts to another device on the same chip without local gates.
}\end{figure}

Fig.~\ref{fig-supp-device} shows an optical microscope image of the device used for the study.
The launching contact, connected to terminal 2, can be seen on the left side.
It has been fabricated according to the recipe described in Ref.~\onlinecite{Wang2013}.
The optical illumination is from the left under an angle of $\sim 45^\circ$, with polarization perpendicular to the edge of the launching contact. 
It is focused with a parabolic mirror onto the atomic force microscope probe tip.

\section{Equivalent circuit model}

\begin{figure}[t]
\centering
\begin{overpic}{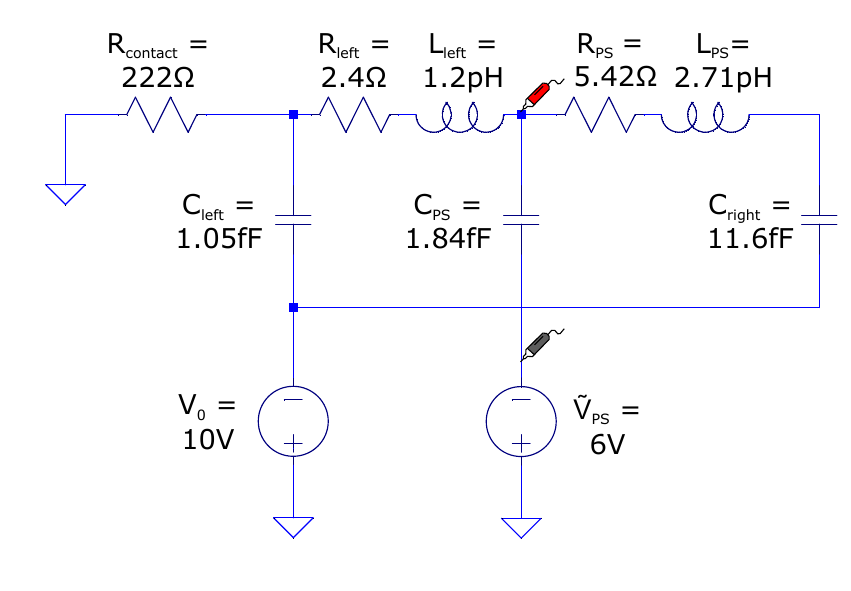}
\put(0,65){\textbf{\textsf{a}}}
\end{overpic}
\begin{overpic}{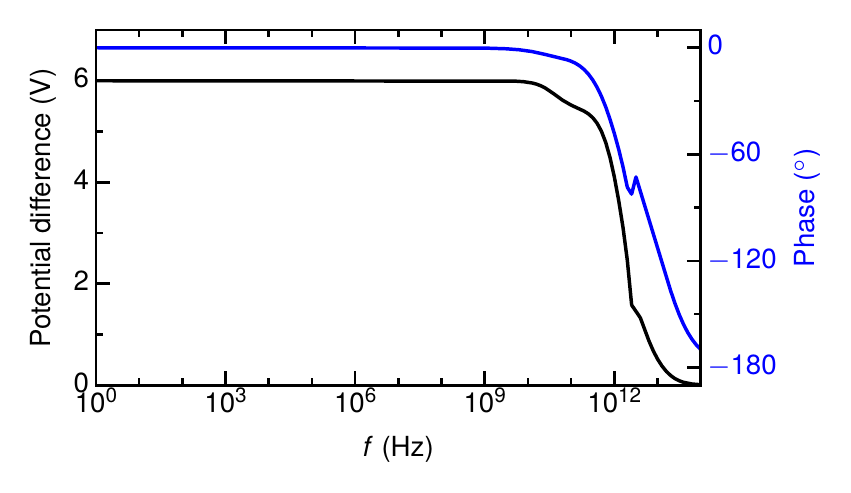}
\put(0,55){\textbf{\textsf{b}}}
\end{overpic}
\caption{
\label{fig-supp-equivalent-circuit}
\textbf{Equivalent circuit model of the phase modulator.}
\textbf{a} Scheme of the equivalent circuit model with definitions of all the parameters.
The potential difference is measured between the two voltage probes indicated in red and grey.
\textbf{b} Bode plot of the frequency response of the potential difference between the voltage source and the phase shifter gate.
}\end{figure}

We estimate the frequency response by modelling our device with an equivalent circuit.
The software used is LTspice IV 4.23.
In Fig.~\ref{fig-supp-equivalent-circuit}a we show the model.
The contact resistance $R_\mathrm{contact}$ is a relevant parameter and should be kept small.
The value of \SI{220}{\ohm} was calculated from a contact resistance of \SI{1000}{\ohm\micro\metre}, which is well within technical capabilities and easily reachable for typical edge contacted graphene devices.\cite{Wang2013}
The assumed channel width was \SI{4.5}{\micro\metre}.
In order to calculate the resistances and inductances in Fig.~\ref{fig-supp-equivalent-circuit}a we used the Drude conductivity of graphene with a carrier lifetime of $500$\,fs, and a relative permittivity of h-BN of 3.56.
The simulated length of the left, center, and right gate is 350\,nm, 350\,nm, and \SI{2.2}{\micro\metre} respectively.
The voltage applied to the left and right gate are 10\,V and 6\,V to the phase shifter gate in order to simulate a phase shift of $\pi$.

In Fig.~\ref{fig-supp-equivalent-circuit}b we see the frequency response of the system.
The frequency cutoff of this equivalent circuit model is above $\sim 1$\,THz for these device parameters.

\section{Measurement details}

\begin{figure*}[t]
\centering
\includegraphics{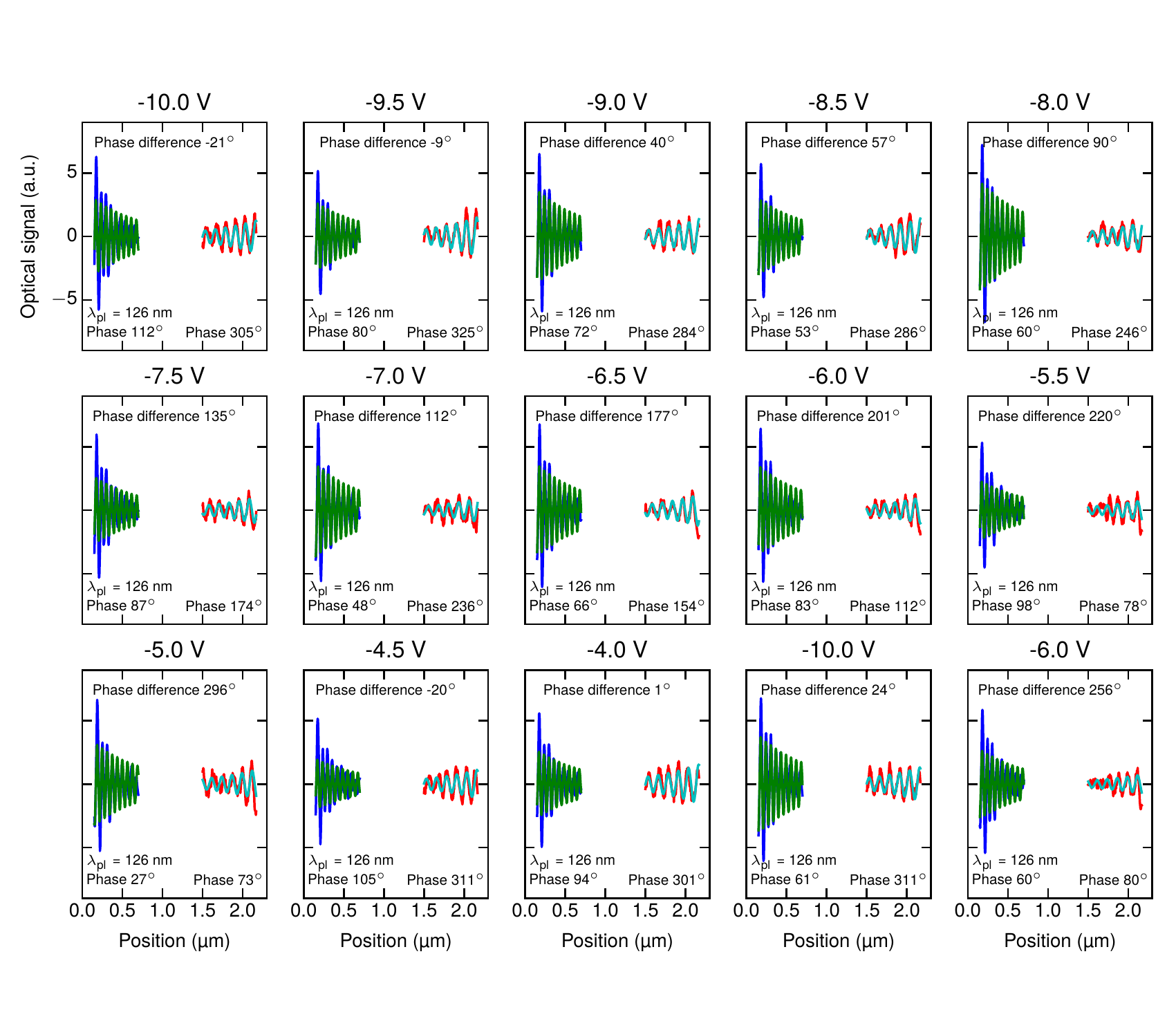}
\caption{
\label{fig-supp-fits}
\textbf{Experimentally measured plasmon signals.}
 Measured edge-reflected (blue line) and pad-launched (red line) plasmon signals, and their respective best-fits (green and light blue lines) for different values of the phase-shifter gate voltage between $-10$\,V and $-4$\,V. 
All the other gates are held at $-10$\,V. 
For the sake of clarity the two signals are shown closer in space than they actually are.  
}\end{figure*}

In Fig.~\ref{fig-supp-fits} we show the raw data of the launched as well as edge reflected plasmons and the corresponding fits.
It is clear that the fits and data agree well.

As explained in the main text we fit both reflected and launched fringes using $\Re(ae^{i\phi}e^{ik x})$, where $\phi$ is the phase we are interested in and $a$ is a real valued amplitude and we keep the plasmon wavevector $k=\kp + k_\mathrm{i}i = 2\pi/\SI{126}{\nano\metre} - \SI{1.5i}{\micro\metre}^{-1}$ constant for all fits.
This guarantees that the phase of the fringes is fully captured in $\phi$.
The position of the graphene edge is measured from the simultaneously acquired topography and is used as reference point.
This enables us to use the phase of the edge reflected fringes as an absolute reference to compensate for potential drifts in the length of one of the interferometric arms which would lead to an observed change in fringe position in absence of an actual change of the plasmon phase.

\section{Topography}

\begin{figure}[t]
\centering
\includegraphics{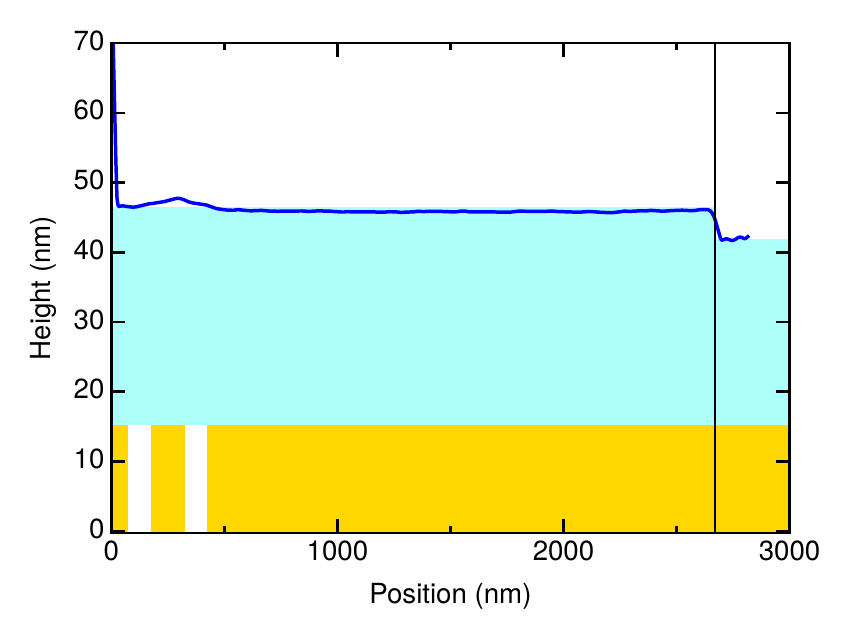}
\caption{
\label{fig-supp-afm-topography}
\textbf{Topography of the device.}
Height profile of the device shows an extremely flat device for the whole length of the device.
On the very left the beginning of the launching contact is visible. 
}\end{figure}

Fig.~\ref{fig-supp-afm-topography} shows the topography measured simultaneously with the optical signal shown in Fig. 3a of the main text. 
It is clear from the measurement that over a distance of 500\,nm at the phase shifter region the topographic height changes by less than 2\,nm and no sagging or other unevenness of the structure is visible. 
We thus do not include any type of surface effects in our analysis.

\section{Plasmon wavelength in a uniform system}
A  uniform two-dimensional (2D) electron system supports plasmon modes in the form of plane waves with wavevector $\kp$. The latter depends on the illumination frequency $\omega$, $\kp = \kp(\omega)$, in a way that can be found by solving the equation~\cite{Vignale2005}

\begin{equation}\label{eq:plasmonuniform}
1-v(\kp,\omega)\tilde{\chi}_{\rm h}(\kp,\omega)=0~,
\end{equation}
where $v(q,\omega)$ is the effective electron-electron interaction in the plane where electrons move and 
$\tilde{\chi}_{\rm h}(q,\omega)$ is the {\it proper} density-density resonse function of the homogeneous 2D electron system~\cite{Vignale2005}.
Here $q$ is a wavevector space variable and $\kp$ is the specific numeric value assumed when the system supports a plasmon mode at the frequency considered (that is $\kp=2\pi/\SI{128}{\nano\metre}$ in our experiment).

For our encapsulated graphene heterostructures, we write the effective interaction potential as
\begin{equation}\label{eq:eeinteraction}
v(q,\omega)\equiv\frac{2\pi e^2}{q\sqrt{\epsilon_x\epsilon_z}}\mathcal{F}(q,\omega)~,
\end{equation}
where $\epsilon_x$ ($\epsilon_z$) is the frequency-dependent in-plane (out-of-plane) permittivity of hBN, and $\mathcal{F}(q,\omega)$ is a form factor that takes into account finite-thickness effects and the presence of a metal gate. Since our scattering theory needs a real valued electron-electron interaction to be fully consistent, we neglect the small (See Measurements details in the main text) imaginary parts of the dielectric constants $\epsilon_x$ and $\epsilon_z$.

The interaction potential in a  vacuum/hBN/G/hBN/metal heterostructure has been  calculated e.g.~in the Supplementary Information of Ref.~\onlinecite{Alonso-Gonzalez2016}. It reads as following
\begin{widetext}
\begin{equation}
\mathcal{F}(q,\omega)=\frac{\displaystyle \tanh{\left[ q\sqrt{\frac{\epsilon_x}{\epsilon_z}}(d+d^\prime) \right]} + \frac{\sinh{\left[ q\sqrt{\frac{\epsilon_x}{\epsilon_z}}(d - d^\prime) \right]}}{\cosh{\left[ q\sqrt{\frac{\epsilon_x}{\epsilon_z}}(d + d^\prime) \right]}}+ \frac{1}{\sqrt{\epsilon_{x}\epsilon_{z}}}\left\{1-\frac{\cosh{\left[ q\sqrt{\frac{\epsilon_x}{\epsilon_z}}(d^\prime - d) \right]}}{\cosh{\left[ q\sqrt{\frac{\epsilon_x}{\epsilon_z}}(d + d^\prime) \right]}}\right\}}{1 + \frac{1}{\sqrt{\epsilon_{x}\epsilon_{z}}}\tanh{\left[ q\sqrt{\frac{\epsilon_x}{\epsilon_z}}(d + d^\prime) \right]}}~.
\end{equation}
\end{widetext}
Here $d=27~{\rm nm}$ is the distance between the graphene and the bottom gate, while $d'=5~{\rm nm}$ is the thickness of the upper hBN layer.

In the Random Phase Approximation (RPA)~\cite{Vignale2005}, the proper density-density response function is replaced by the non-interacting density-density response function of a 2D massless Dirac fermion system. The latter can be written as 
\begin{equation}\label{eq:chigraphene}
\tilde{\chi}_{\rm h}(q,\omega)=\frac{\bar{E}_{\rm F}q^2}{\pi \hbar^2 \omega^2}\mathcal{G}(q,\omega, \bar{E}_{\rm F})~,
\end{equation}
where $\bar{E}_{\rm F}$ is the Fermi energy of the uniform electron system, and $\mathcal{G}(q,\omega,\bar{E}_{\rm F})$ is given, up to second order in $q$, by
\begin{equation}\label{eq:gdefinition}
\mathcal{G}(q,\omega,\bar{E}_{\rm F})=1-\frac{\hbar^2 \omega^2}{4\bar{E}_{\rm F}^2}+\frac{3}{4}\frac{v_{\rm F}^2 q^2}{\omega^2}~.
\end{equation}
Making use of Eqs.~(\ref{eq:eeinteraction})-(\ref{eq:chigraphene}), we can rewrite Eq.~(\ref{eq:plasmonuniform}) as
\begin{equation}\label{eq:final}
1-\frac{2 e^2 \bar{E}_{\rm F} |\kp|}{\sqrt{\epsilon_x\epsilon_z}\hbar^2 \omega^2}\mathcal{F}(\kp,\omega)\mathcal{G}(\kp,\omega,\bar{E}_{\rm F})=0~.
\end{equation}
This equation must be solved numerically for $\kp$. 

According to electrostatic simulations (see Fig 2a in the main text) the carrier density far away from the phase-shift gate is $-7.29 \times 10^{12}~{\rm cm^{-2}}$. From the solution of Eq.~(\ref{eq:final}), we find $\kp=4.89 \times 10^{-2}~{\rm nm^{-1}}$ and a plasmon wavelength $\lambda_{\rm P}=128~{\rm nm}$.
\section{Lippman-Schwinger-RPA scattering theory for two-dimensional plasmons}
The reflection and transmission coefficients for plasmons impinging on a localized inhomogeneity in the graphene sheet are determined by the proper density-density response function of the {\it inhomogeneous} 2D electron system. The latter depends~\cite{Vignale2005} on {\it two} wavevectors and on the frequency $\omega$.  As stated above, in the RPA, we can replace the proper response function with the  response function of a noninteracting system.
For inhomogenous graphene sheets, even the latter quantity is not known exactly. In this work we use the following approximation for the long-wavelength limit of the non-interacting density-density response function of an inhomogeneous 2D massless Dirac fermion system \cite{TorreTobepublished2016}:
\begin{equation}\label{eq:chinonhomogeneous}
\begin{split}
\tilde{\chi}(\bm q,\bm q',\omega) & =\frac{E_{\rm F}(\bm q-\bm q')\bm q \cdot \bm q'}{\pi S \hbar^2 \omega^2}
-\frac{E_{\rm F}^{-1}(\bm q-\bm q')\bm q \cdot \bm q'}{4 \pi S}\\
& +\frac{3 v_{\rm F}^2 E_{\rm F}(\bm q-\bm q')(\bm q \cdot \bm q')^2}{4\pi S\hbar^2 \omega^4}~.
\end{split}
\end{equation}
Here $S$ is the 2D electron system area, $E_{\rm F}(\bm q)$ is defined by
\begin{equation}
E_{\rm F}(\bm q)=\int d \bm r~e^{-i \bm q \cdot \bm r} \hbar v_{\rm F}~{\rm sgn}[n(\bm r)]\sqrt{\pi |n(\bm r)|}~,
\end{equation}
where $v_{\rm F}$ is the graphene Fermi velocity and $n(\bm r)$ the local carrier density, while
\begin{equation}
E_{\rm F}^{-1}(\bm q)=\int d \bm r~e^{-i \bm q \cdot \bm r} \left[\hbar v_{\rm F}~{\rm sgn}[n(\bm r)]\sqrt{\pi |n(\bm r)|}\right]^{-1}~.
\end{equation}
Eq.~(\ref{eq:chinonhomogeneous}) is valid up to $\mathcal{O}(q^4)$. 

The above formula generalizes Eq.~(\ref{eq:chigraphene}), by allowing spatial variations of the Fermi energy. Eq.~(\ref{eq:chinonhomogeneous}) reduces to Eq.~(\ref{eq:chigraphene}) in the case of a  uniform Fermi energy.

Let's now write the local Fermi energy as the sum of an asymptotic value $\bar{E}_{\rm F}$ plus a perturbation $\delta E_{\rm F}(x)$, which vanishes far away from the phase-shift gate and depends only on $x$. The latter represents the carrier density variation induced by the presence of the phase-shift gate.

The proper density-density response function can be decomposed according to
\begin{equation}
\tilde{\chi}(\bm q,\bm q',\omega)=\delta_{\bm q \bm q'}\tilde{\chi}_{\rm h}(q,\omega)+\frac{1}{L_x}\delta_{q_yq_y'}\delta \tilde{\chi}( q_x, q_x',q_y,\omega)~,
\end{equation}
where $\tilde{\chi}_{\rm h}(q,\omega)$ is calculated from (\ref{eq:chigraphene}), and $L_x$ is the length of the 2D electron system in the $x$ direction. 

The equation describing plasmons in an inhomogeneous system is\cite{TorreTobepublished2016}
\begin{equation}\label{eq:plasmoninhomogeneous}
\sum_{\bm q'}\left[\delta_{\bm q \bm q'}- v(q,\omega)\tilde{\chi}(\bm q,\bm q',\omega)\right]\Phi(\bm q',\omega)=0.
\end{equation}
Here $\Phi(\bm q,\omega)$ is the self-consistent Hartree potential on the graphene sheet.

To calculate reflection and transmission coefficients for plasmons impinging on the region of the phase-shift gate, we need a solution of Eq.~(\ref{eq:plasmoninhomogeneous}) with asymptotic behavior
\begin{equation}\label{eq:asymptoticbehaviour1d}
\Phi(x,\omega)\simeq 
\begin{cases}
e^{i  \kp(\omega) x}+r_{\omega}e^{-i  \kp(\omega) x}\quad x \to -\infty\\
t_{\omega}e^{i \kp(\omega) x}\quad x \to +\infty
\end{cases}~.
\end{equation}

As in the theory of quantum-mechanical scattering, we enforce this asymptotic condition by following the usual Lippmann-Schwinger steps\cite{Sakurai1994}. We first introduce the transition matrix
\begin{equation}\label{eq:tmatrixdefinition1d}
T(q_x,\omega)\equiv \frac{1}{L_x}\sum_{q_x'} \delta \tilde{\chi}(q_x, q_x',q_y=0,\omega)\Phi(q_x',\omega)~,
\end{equation}
which satisfies the following equation 
\begin{equation}\label{eq:tmatrixequation1d}
\begin{split}
&T(q_x,\omega)  =\delta \tilde{\chi}(q_x, \kp(\omega),q_y=0,\omega)\\
&+\frac{1}{L_x} \sum_{ q_x'} \delta \tilde{\chi}(q_x, q_x',q_y=0,\omega)W_{\rm h}^{(+)}\left(|q_x'|,\omega\right)T(q_x',\omega)~.
\end{split}
\end{equation}
In Eq.~(\ref{eq:tmatrixequation1d}), the screened interaction is\cite{TorreTobepublished2016}
\begin{equation}\label{eq:whom+}
W_{\rm h}^{(+)}(q,\omega)=\frac{1}{v^{-1}(q,\omega)-\tilde{\chi}_{\rm h}(q,\omega)+i0^+}~.
\end{equation}

The transition matrix is directly related to the transmission and reflection coefficients by
\begin{align}
t_{\omega}=1-iC(\omega)T(\kp(\omega),\omega)~,\\
r_{\omega}=-iC(\omega)T(-\kp(\omega),\omega)~,
\end{align}
where
\begin{equation}\label{eq:resdefinition}
C(\omega)
=\frac{2 \pi e^2\mathcal{F}(\kp,\omega)}{\sqrt{\epsilon_x \epsilon_z}\left[1+\kp\frac{\mathcal{F}'(\kp,\omega)}{\mathcal{F}(\kp,\omega)}+\kp\frac{\mathcal{G}'(\kp,\omega)}{\mathcal{G}(\kp,\omega)}\right]}~.
\end{equation}
Here the prime stands for derivation with respect to the wavevector argument.

Eq.~(\ref{eq:tmatrixequation1d}) is the most important result of this Section, and has been  used to calculate the phase of the transmitted plasmon. Indeed, the theoretical curve shown in Fig.~4 of the main text has been obtained from a direct numerical solution of Eq.~(\ref{eq:tmatrixequation1d}), using a first-order finite-element method.

In the following, we introduce two methods to find approximate solutions of Eq.~(\ref{eq:tmatrixequation1d}). These provide a good qualitative understanding of the scattering problem and can be useful tools in designing new experiments.

\subsection{First-Order Born Approximation}
If the density perturbation is very small with respect to the equilibrium density, we can use perturbation theory in powers of $\delta\tilde{\chi}$.
The result of first-order perturbation theory is known in the the context of scattering theory as First-Order Born Approximation~(FOBA) \cite{Garcia-Pomar2013}. It is obtained by neglecting the term in the second line of Eq.~(\ref{eq:tmatrixequation1d}).
This yields simple expressions for the scattering coefficients:
\begin{align}
t_{\omega}^{\rm FOBA}=1-iC(\omega)\delta\tilde{\chi}(\kp(\omega),\kp(\omega),0,\omega)~,\\
r_{\omega}^{\rm FOBA}=-iC(\omega)\delta\tilde{\chi}(-\kp(\omega),\kp(\omega),0,\omega)~.
\end{align}
We notice that the FOBA is unable to explain phase shifts exceeding $\pi/2$ and can be used only in a small range of phase-shift gate  voltages. This is because, in the FOBA, the phase of the transmission coefficient is
\begin{equation}
\arg\left(t_\omega^{\rm FOBA}\right)=-\arctan\left[C(\omega)\delta\tilde{\chi}(\kp(\omega),\kp(\omega),0,\omega)\right]~.
\end{equation}
\subsection{Optical Path Length Approximation}
Here we provide a rigorous derivation of the Optical Path Length Approximation (OPLA) introduced in the main text. This becomes exact when the local refractive index varies slowly on the length-scale of the plasmon wavelength. Contrary to the FOBA, the  optical path length approximation does not rely on the smallness of $\delta \tilde{\chi}$.

For plasmons propagating in the ${\hat{\bm x}}$ direction, the general plasmon equation (\ref{eq:plasmoninhomogeneous}) can be rewritten in real space as following:
\begin{equation}\label{eq:scatteringrealspace}
\begin{split}
\int dx' \left[v^{-1} (x-x',\omega)-\tilde{\chi}(x,x',q_y=0,\omega) \right] \Phi(x',\omega)=0 ~,
\end{split}
\end{equation}
where
\begin{equation}\label{eq:inversepotential}
v^{-1}(x,\omega)=\int \frac{dq}{2\pi} \frac{e^{iqx}}{v(q,\omega)}~.
\end{equation}
Introducing the eikonal ansatz
\begin{equation}\label{eq:eikonalansatz}
\Phi(x,\omega)=\exp[i \kp S(x)]~,
\end{equation}
we derive an equation for the function $S(x)$:
\begin{equation}\label{eq:eikonalequation}
\begin{split}
&\int dx' v^{-1}(x-x',\omega)e^{-i k_{\rm pl}[ S(x)-S(x')]}=\\
&=\left[\frac{E_{\rm F}(x)}{\pi \hbar^2 \omega^2}-\frac{1}{4\pi E_{\rm F}(x)}\right]k_{\rm pl}^2\left[S'(x)\right]^2\\
&+\frac{3v_{\rm F}^2E_{\rm F}(x)}{4\pi \hbar^2 \omega^4 }k_{\rm pl}^4\left[S'(x)\right]^4+\mathcal{O}\left(k_{\rm pl}\right)~.
\end{split}
\end{equation}

To leading order in $\kp$, the left hand side of the previous equation can be approximated by
\begin{equation}\label{eq:integraloperator}
\frac{1}{v(k_{\rm pl}S'(x),\omega)}~.
\end{equation}
We define the {\it local} plasmon wavevector $\kp(x)$ as the solution of 
\begin{equation}\label{eq:localplasmonwavevector}
1-\frac{2 e^2 E_{\rm F}(x) |\kp(x)|}{\sqrt{\epsilon_x \epsilon_z}\hbar^2 \omega^2}\mathcal{F}(\kp(x),\omega)\mathcal{G}(\kp(x),\omega,E_{\rm F}(x))=0~.
\end{equation}
The local wavevector $\kp(x)$ reduces to $\kp$ for $x$ far away from the phase-shifter gate.
Keeping only leading terms in the limit $\kp \to \infty$ we can simplify Eq.~(\ref{eq:eikonalequation}) to
\begin{equation}\label{eq:wkbapprox}
\kp S'(x)=\kp(x)~.
\end{equation}
The corresponding solution for the potential is
\begin{equation}\label{eq:eikonalsolution}
\Phi(x,\omega) \propto \exp\left[i\int_{0}^x dx' \kp(x')\right]~,
\end{equation}
and the phase of the scattering coefficient is therefore
\begin{equation}\label{eq:argtwkb}
\arg(t_\omega)=\int_{-\infty}^{\infty} dx' \left[ \kp(x')-\kp\right]~.
\end{equation}
The above result has been used to calculate approximatively the phase of the transmitted plasmon. The result is shown in Fig.~4 of the main text as a blue curve.


\end{document}